\documentclass[reprint,superscriptaddress,amsmath,amssymb,aps,prx,longbibliography]{revtex4-2}

\usepackage{graphicx} 
\usepackage{setspace}

\begin{document} 

\title{Quantifying and controlling prethermal nonergodicity in interacting Floquet matter}

\author{K.~Singh}
\author{C.~J.~Fujiwara}
\author{Z.~A.~Geiger}
\author{E.~Q.~Simmons} 
\author{M.~Lipatov}
\author{A.~Cao} 
\author{P.~Dotti} 
\author{S.~V.~Rajagopal} 
\author{R.~Senaratne} 
\author{T.~Shimasaki}
\affiliation{Department of Physics, University of California, Santa Barbara, California 93106, USA}
\author{M.~Heyl} 
\author{A.~Eckardt}
\affiliation{Max-Planck-Institut f\"ur Physik komplexer Systeme, N\"othnitzer Str.\ 38, 01187 Dresden, Germany} 
\author{D.~M.~Weld}
\email{weld@ucsb.edu}
\affiliation{Department of Physics, University of California, Santa Barbara, California 93106, USA}

\begin{abstract} 
The use of periodic driving for synthesizing many-body quantum states depends crucially on the existence of a prethermal regime, which exhibits drive-tunable properties while forestalling the effects of heating. This motivates the search for direct experimental probes of the underlying localized nonergodic nature of the wave function in this metastable regime. We report experiments on a many-body Floquet system consisting of atoms in an optical lattice subjected to ultrastrong sign-changing amplitude modulation. Using a double-quench protocol we measure an inverse participation ratio quantifying the degree of prethermal localization as a function of tunable drive parameters and interactions. We obtain a complete prethermal map of the drive-dependent properties of Floquet matter spanning four square decades of parameter space. Following the full time evolution, we observe sequential formation of two prethermal plateaux, interaction-driven ergodicity, and strongly frequency-dependent dynamics of long-time thermalization.  The quantitative characterization of the prethermal Floquet matter realized in these experiments, along with the demonstration of control of its properties by variation of drive parameters and interactions, opens a new frontier for probing far-from-equilibrium quantum statistical mechanics  and new possibilities for dynamical quantum engineering.
\end{abstract}

\maketitle

Time-periodic driving is a powerful technique for synthesizing tailored quantum matter with drive-dependent properties and phase structures beyond the constraints imposed by thermodynamic equilibrium~\cite{Eckardt:2017aa,Holthaus,EsslingerHaldane,BlochChern,Rechtsman:2013aa,chinFMshakenlattice,Zhang:2017aa,Choi:2017aa, Moessner,Roushan:2016aa,Flaschner:2016aa,miyake-Harper, WeldFloquet}, and therefore beyond a description in terms of thermal Gibbs ensembles. 
In many cases of interest, driven systems might still localize in metastable prethermal states~\cite{Abanin:2015aa,KUWAHARA201696,huveneers,chetanNorm,Ueda} as a consequence of approximate integrals of motion, given by a high-frequency approximation to the effective Floquet Hamitonian or, for weak interactions, by a macroscopic number of operators describing the occupations of single-particle Floquet states. Such prethermal states are conjectured to be described by a periodic Gibbs ensemble~\cite{Lazarides:2014aa,Lazarides:2014ab,chandranfloquet,Weidinger:2017aa,KUWAHARA201696,EcksteinPGE,KnapSteadyStates}. While pioneering experiments have explored the uses and properties of metastable driven quantum systems (mainly in the regime of the high-frequency approximation)~\cite{BlochHeating,Messer,1808.07462,interband_eckardt_sengstock_simonet,Porto2019,sengstock-dqpt,BlochMB}, directly probing the localized nature of the prethermal many-body quantum state has remained an outstanding challenge.

In this work, we experimentally probe a universal quantitative measure for non-ergodic localization of a driven quantum system based on a stroboscopically time-averaged return probability to an initial energy-localized state. This quantity directly corresponds to an inverse participation ratio quantifying  localization in the eigenbasis of the approximate Floquet Hamiltonian, and can be experimentally measured using a double-quench protocol starting from an almost fully Bose-condensed undriven ground state. The experiments we report use tunably-interacting $^{7}$Li atoms in an amplitude-modulated optical lattice to create and probe prethermal Floquet matter. Uniquely, our experiment enables the use of amplitude modulation extending from 10 to 1000 percent of the static lattice depth and drive frequencies extending from 0.1 to 10 times the lattice band gap. This regime of ultrastrong drive amplitudes is previously unrealized experimentally, and this range of drive frequencies extends through and well beyond typical ``high-frequency'' regimes of Floquet engineering. Realization of these extreme parameter values allows us to map out a sharp threshold between differently-localized regimes of prethermal ergodicity breaking. Observed characteristics of the prethermal state quantitatively confirm theoretical predictions based on a periodic Gibbs ensemble. Tracking the time evolution of the driven system we observe and quantify not only the formation of a prethermal non-ergodic plateau, but also the long-time departure from it, either by a transition to a second prethermal plateau or, for stronger interactions, by the onset of ergodicity.

The initial condition for all experiments discussed here is a Bose-condensed gas of lithium loaded into the ground band of a static 1D optical lattice of depth $V_{0}$ and wavelength $\lambda=1064\ \mathrm{nm}$ at zero quasimomentum.  Interatomic interactions are set to the desired value using a magnetic Feshbach resonance.  The system is quenched into the Floquet Hamiltonian by applying lattice amplitude modulation at some frequency $\omega$ and relative amplitude $\alpha$, keeping the cycle-averaged lattice depth fixed at $V_0$. After some modulation time, the atoms are quenched back to the original  optical lattice, bandmapped, and imaged in order to measure the resulting distribution in the eigenbasis of the undriven lattice.  This double-quench protocol provides a direct probe of the evolution of system properties under the Floquet Hamiltonian, and of localization in the Floquet state basis. It is not restricted to our particular experimental setup, and therefore provides a general strategy for measuring ergodicity breaking that can be useful also in other experimental contexts. More details on the experimental protocol appear in the appendices.

If interactions are tuned to zero, the Hamiltonian of this driven system is \begin{equation}
H(t) = -(\hbar^{2}/2m)\partial_{x}^{2} + V_{0}[1+\alpha \sin(\omega t)] \cos^{2}(k_{L}x),
\end{equation}
where $x$ is position along the lattice,  $m$ is the atomic mass, $k_{L}=2\pi/\lambda$ is the wavenumber of the lattice laser, and $V_0=10 E_{R}$ is the static lattice depth, with recoil energy $E_{R}= \hbar^{2}k_{L}^{2}/2m$. It is intuitively useful to note that this is a quantum mechanical version of the Hamiltonian for a rigid pendulum with a vertically modulated pivot point, with the compact phase variable of the pendulum replaced by the position $x$.  The role of classical rigid pendula as archetypes for the study of stability and instability in driven systems provides natural motivation for the quantum mechanical experiments reported here.   

The drive frequency can be expressed dimensionlessly as $\Omega=\omega/\omega_{0}$ where  $\omega_0=2\sqrt{V_{0}E_{R}}/\hbar$ is the frequency  of harmonic motion in a single lattice site, which approximates the lowest band gap of the static lattice. Our experiments explore four square decades of drive parameter space, with a dimensionless frequency from $\Omega=0.1$ to $\Omega=10$ and a dimensionless amplitude from $\alpha=0.1$ to $\alpha=10$. This wide range represents a challenge  both for theory, due to the absence of any reliably small scale in the problem, and for experiment, due to the difficulty of attaining modulation amplitudes greater than 100\%. We realize the strongly-driven regime $\alpha>1$ by simultaneously modulating two orthogonally-polarized co-axial 1D optical lattices with a relative spatial phase shift of $\lambda/4$. Exploration of this ultrastrong driving regime is a key experimental novelty of this work, and it is required for access to the majority of the parameter space we explore in maps of prethermal Floquet matter properties like those shown in  Fig.~1.  


We characterize and quantify prethermalization in our Floquet system via the fraction $f_0$ of atoms which occupy the single-particle ground state after the modulation is quenched off. This quantity naturally contains information about heating in the Floquet system since those atoms which are excited out of the initial ground state automatically lead to a reduced $f_0$. Beyond this intuitive argument, we find that $f_0$ in fact represents a powerful quantitative measure for localization and therefore for prethermalization and the absence of heating: specifically, $f_0$ can be directly related to an inverse participation ratio ($\mathrm{IPR}$) in our experimental context. An $\mathrm{IPR}$, defined as $\mathrm{IPR} = \sum_n |\langle \psi_0 | n \rangle |^4$, quantifies how strongly the state $|\psi_0\rangle$, here representing the initial condition, is localized in the basis $|n \rangle$, which in our case is the eigenbasis of the Floquet time-evolution operator. The participation ratio $1/\mathrm{IPR}$ measures the number of Floquet states $|n \rangle$ required to represent the initial state. In a localized prethermal state the $\mathrm{IPR}$ takes on a nonzero value whereas in the delocalized ergodic regime the $\mathrm{IPR}$ becomes vanishingly small. We find, crucially, that if interactions are neglected the long-time average of $f_0$ measured stroboscopically at integer multiples of the driving period, $\bar{f}_0$, is exactly equivalent to the IPR; this is demonstrated in detail in appendix~\ref{IPRappendix}. While the $\mathrm{IPR}$ is a standard diagnostic of localization, the experimental measurement of such a quantity is, in general, very challenging for a many-body system. For an interacting system in the prethermal regime described by the PGE, the exact identification of $f_0$ as an IPR does not hold, since for example the condensate fraction will be depleted slightly already in the ground state by quantum fluctuations, but $f_0$ remains a useful and experimentally accessible metric for characterizing the properties of the interacting driven system. In particular, a non-zero $f_0$ still indicates a non-ergodic interacting prethermal state. We note that the IPR-based interpretation of $\bar{f}_0$ as a quantitative measure for prethermal ergodicity breaking proposed here applies to a wide range of bosonic quantum systems, since it essentially relies only on a small quantum depletion of the pre-quench condensate.

\begin{figure*}[tbh]
\includegraphics[width=0.85\linewidth]{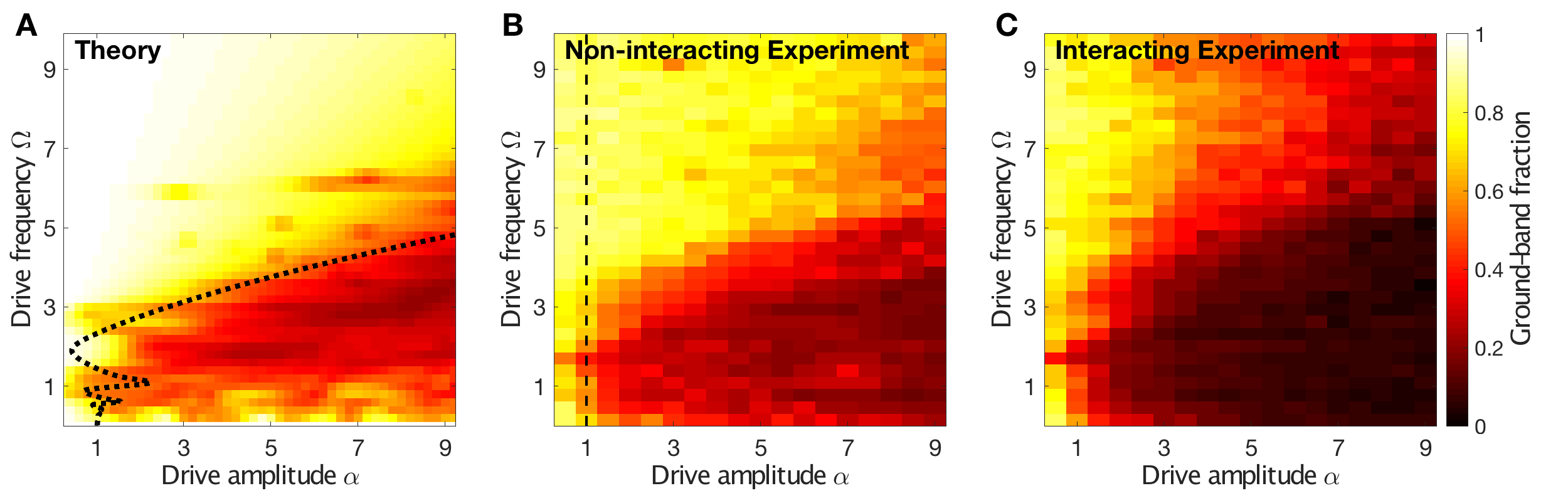}
\caption{Mapping the prethermal state in the space of drive parameters.  {\bf A:} Theoretical map of the projected static ground band occupation fraction for a non-interacting periodic Gibbs ensemble, as a function of normalized drive frequency $\Omega$ and normalized drive amplitude $\alpha$. Dotted line shows classical stability boundary of the equivalent driven pendulum. {\bf B:} Experimental measurement of normalized population in the center of the lowest band after a 500~$\mu$s hold of a non-interacting quantum gas in a modulated optical lattice, as a function of the same parameters. The $\alpha>1$ region to the right of the vertical dashed line is inaccessible without the sign-changing amplitude modulation introduced in this work. {\bf C:} Same measurement as B, but in the presence of interatomic interactions ({\it s}-wave scattering length 30~nm). Colorbar and axes are the same for all three panels.}
\label{fig:PhaseDiagram1} \label{fig1}
\end{figure*}

\begin{figure*}[tbh]
\centering
\includegraphics[width=0.85\linewidth]{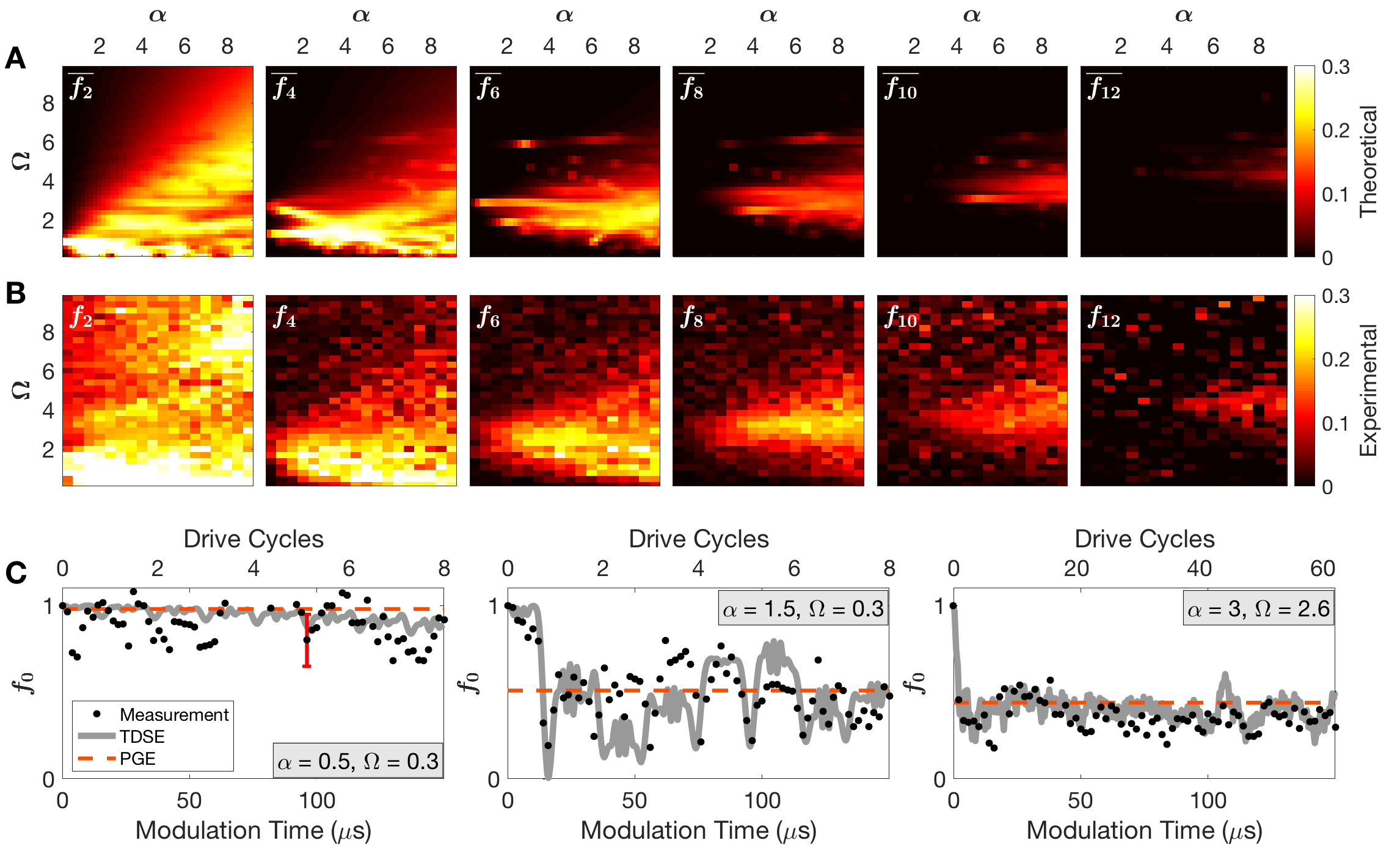}
\caption{Characterizing the prethermal periodic Gibbs ensemble.  \textbf{A:} PGE prediction for occupation of excited even bands as a function of drive parameters $\alpha$ and $\Omega$. \textbf{B:} Measured atom number fraction in the central 40\% of the same bands.  All axes and colorbars are the same as for the theory panels.  \textbf{C:} Measured  $f_0$ versus modulation time for the first 150~$\mu$s of the drive, for three different values of $(\alpha,\Omega)$ all for vanishing interactions. Dashed line shows PGE prediction, solid line shows TDSE prediction of the fluctuating evolution of a spatially homogeneous non-interacting system. Each point represents a single run; error bar shows representative estimated fractional error due to shot-to-shot number fluctuation. For these data, the optical lattice is immediately quenched back to the initial 10 $E_{R}$ static lattice at the indicated modulation time and then band-mapped.}
\label{fig:ExcitedAtoms}
\end{figure*}

The dependence of Floquet matter attributes on drive properties can be directly calculated in the non-interacting case. We calculate $\bar{f}_0(\alpha,\Omega)$ from the periodic Gibbs ensemble $\rho\propto\exp(-\sum_i\eta_i\hat{n}_i(t))$ characterized by integrals of motion given by the occupations $\hat{n}_i(t)$ of the single-particle Floquet modes $|i(t)\rangle$ of the multi-band Hamiltonian. Here, the coefficients are $\eta_i = \log\big(1+\langle \psi_0|\hat{n}_i(0)|\psi_0\rangle^{-1}\big)$, where the mean-occupations  $\langle \psi_0|\hat{n}_i(0)|\psi_0\rangle=N|\langle 0|i(0)\rangle|^2$ are directly given by single-particle overlaps with the undriven (single-particle) ground state $|0\rangle$. The predicted dependence $\bar{f}_0(\alpha,\Omega)$ is shown in Fig.~1A as a function of dimensionless driving amplitude $\alpha$ and frequency $\Omega$. The theory predicts that $\bar{f}_0\simeq 1$ for a large region in $(\alpha,\Omega)$ space, but as $\Omega$ decreases from large values for any given drive amplitude, there is always some $\alpha$-dependent drive frequency below which $f_0$ sharply decreases to a lower but still non-ergodic value between 0 and 1. Intriguingly, this sharp crossover corresponds approximately to the stable-unstable crossover of the corresponding classical system: a rigid pendulum with a vertically modulated pivot point.  The dashed line in Fig.~1A shows the classical boundary of stability for the downward-pointing pendulum; exploration of a possible quantum analogue of the upward-pointing Kapitza state~\cite{kapitza,Gilary_2003} is an interesting potential direction for future work.

The wide tunability afforded by our flexible experimental platform enables measurement of the full prethermal map of Floquet material properties predicted in Fig.~1A, and direct observation of the Floquet delocalization crossover.  Figures~1B and 1C show experimentally measured maps of $f_{0}(\alpha,\Omega)$ after 500 $\mu$s for non-interacting and interacting samples respectively.  Because experimental quasimomentum resolution is limited by the finite initial spatial size of the condensate and a finite time-of-flight, the experimental maps are based on integrals over the central 40\% of the Brillouin zone.  The non-interacting measurement of Fig.~1B shows good agreement with the PGE-based theoretical prediction of Fig.~1A, thus experimentally confirming the conjectured applicability of the PGE as a model for strongly-driven systems. For these measurements, theory-experiment discrepancies can arise from experimental imperfections or from the fact that the theory plot represents the predicted long-time average occupation over the stroboscopic dynamics as computed from the PGE whereas the experimental data is a snapshot of the occupation at a fixed point in time. The interacting data in Fig.~1C display a qualitatively similar but not quantitatively identical behavior to those in Fig.~1B; the crucial effects of interactions are explored further in the time-evolution measurements discussed below. Both measurements clearly show the predicted amplitude-dependent Floquet delocalization crossover.
 

Measuring higher-band observables in addition to $f_0$ allows a fuller comparison between experimental and theoretical descriptions of the prethermal state. Over the timescales shown in Fig.~2 the dynamics of the noninteracting driven system mainly redistribute population among the lowest few even Bloch bands, as expected from parity conservation at $k=0$. Fig.~2A shows the results of a PGE-based calculation of fractional projected occupations $f_{2\nu}, \nu\!\in\!(1, \ldots 6)$ of the first six even excited bands, as a function of drive parameters $\alpha$ and $\Omega$.  The theory predicts a rich dependence of Floquet material properties on drive parameters, with a distinct map for each projected band population. Our experiment can quantitatively test these predictions by directly imaging such maps. Fig.~2B shows the experimentally measured fractional population of the first six even-parity excited bands at each $(\alpha,\Omega)$ point, using the same double-quench protocol used to produce Fig.~1. The observed close match between theory and experiment lends further support to the PGE-based theoretical description of the prethermal state.  We do not observe significant occupation above the twelfth band, in agreement with the PGE model. These detailed experimental maps of the properties of prethermal Floquet matter throughout a wide range of drive parameter space reveal the intricate structure of the amplitude-dependent Floquet delocalization crossover in the strong-driving regime and constitute the first main result of this report.

Moving beyond such fixed-time maps, it is possible to experimentally explore the full time evolution of $f_0$ and higher-band observables at any point in drive parameter space. This allows direct measurement of the temporal emergence of the prethermal state, as well as investigation of its long-term fate.  As shown in Fig.~2C, the measured time evolution of $f_0$ for various drive parameters shows remarkably rapid attainment of an average value in close agreement with the PGE theoretical prediction, on timescales near a single drive period~\cite{Lazarides:2014aa}. For a drive with $\Omega = 0.3$ and $\alpha = 0.5$, $f_0$ remains close to 1, while for higher drive amplitudes and frequencies it fluctuates around a lower average value which agrees well with the PGE prediction. Since the inverse of the $\mathrm{IPR}$ provides a measure for the number of eigenstates of the time-evolution operator the system can access, it might appear surprising that even at large driving strengths the predicted and measured $f_0$ remains so high throughout the dynamics; as discussed in appendix~\ref{finitenum} this can be understood as a consequence of the eventual dominance of higher-band kinetic energy splittings over any fixed coupling matrix element. 

\begin{figure}
\centering
\includegraphics[width=0.9\linewidth]{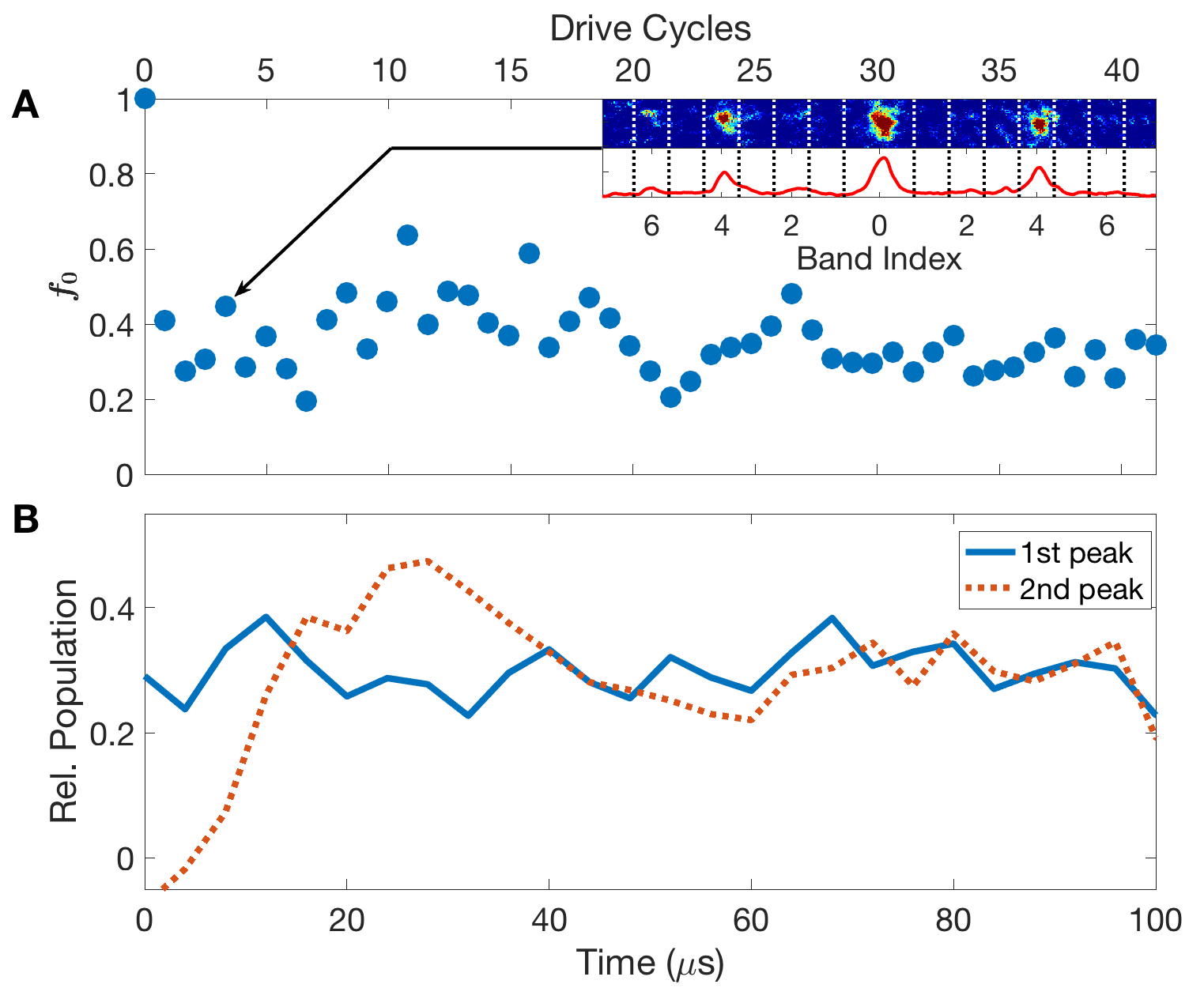}
\caption{Entering the prethermal state. {\bf A:} Normalized ground band occupation $f_0$ as a function of time during modulation with $\alpha\!=\!3$, $\Omega\!=\!2.6$, $a\!=\!0$. The system attains the prethermal value of $f_0$ on time scales comparable to a single drive cycle. These data are reproduced from the rightmost panel of Fig.~2C to facilitate direct comparison with Fig.~3B. Inset shows a sample bandmapped image and its column integration for the indicated data point, with Brillouin zone boundaries and band indices labeled.  {\bf B:} Population fraction in diffracted peaks after diabatic lattice snapoff, as a function of time during the same drive. $x$~axes are the same for the two plots. Gradual attainment of the prethermal steady-state is apparent in the settling of the second-peak population. 
}
\label{fig:timetraceNonInter}
\end{figure}

\begin{figure*}[bt]
\centering
\includegraphics[width= \linewidth]{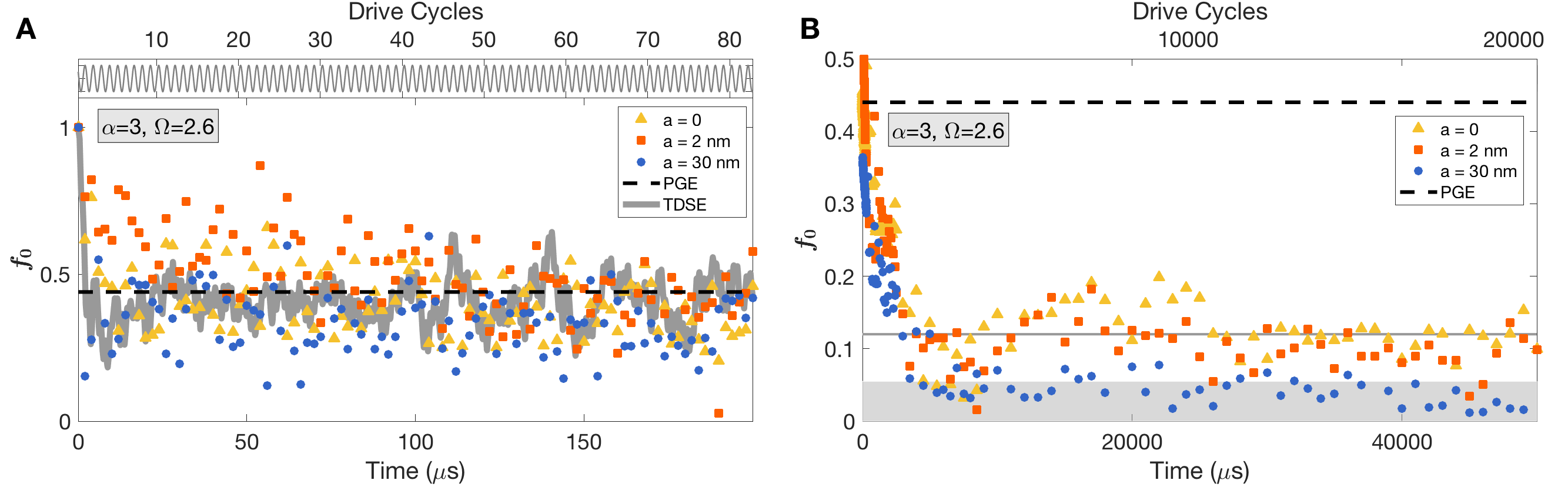}
\caption{Effect of interaction on the long-time evolution of Floquet matter. \textbf{A:} Short-time evolution of $f_0$ for a drive with $\Omega = 2.6$ and $\alpha = 3$, at three different values of the s-wave scattering length $a$. Regardless of interaction strength, $f_0$ fluctuates near the PGE value (dashed line). Solid gray line shows the solution to the time-dependent Schr\"odinger equation for the non-interacting case. Top panel shows the driving waveform. \textbf{B:} $f_0$ as a function of time for much longer times, for the same drive and interaction parameters.  Data are boxcar-averaged into 200~${\mu}s$ bins. The non-interacting and weakly-interacting samples attain a second plateau (solid line) well below the PGE value (dashed line), while the strongest-interacting sample decays to a high-temperature state consistent with ergodicity. Shaded area indicates the estimated noise floor: at long times we measure no significant ground-band occupation only for the strongest-interacting sample. 
}
\label{fig:timetrace}
\end{figure*}

Return probabilities such as $f_0$ yield on general grounds fluctuations of the same order as the mean for unitary dynamics; thus, the fact that we not only measure the expected temporal mean but also observe large temporal fluctuations can be interpreted as a signature of the unitary quantum character of the real-time dynamics of the experiment.  In principle, such fluctuations also contain information on the spectrum of the Floquet Hamiltonian.  To enable qualitative comparison to the expected form of these fluctuations, the solid lines in Fig.~2C show the result of time-dependent Schr\"odinger equation integration for a non-interacting perfectly spatially homogeneous sample. While we do not expect the spatially-inhomogeneous experiment to perfectly reproduce this very simple theory, we do observe good qualitative agreement for the mean value and typical fluctuation amplitude, supporting the picture of these fluctuations as a signature of unitary dynamics.

The rapid but not instantaneous nature of the dephasing responsible for the emergence of the prethermal plateau can be revealed by measuring a different observable: the interference patterns after the system is quenched back to a static lattice which is snapped off immediately rather than band-mapped.  Fig.~3 shows that over the course of a few dozen drive cycles, the occupations of the first two interference peaks approach their quasi-steady-state values.  We interpret this as a signature of the complex drive-dependent dephasing processes which create the prethermal state, and note that the timescale of the measured relaxation lies between the static lattice tunneling time and the inverse of typical static interband energy scales.


\begin{figure*}
\centering
\includegraphics[width= \linewidth]{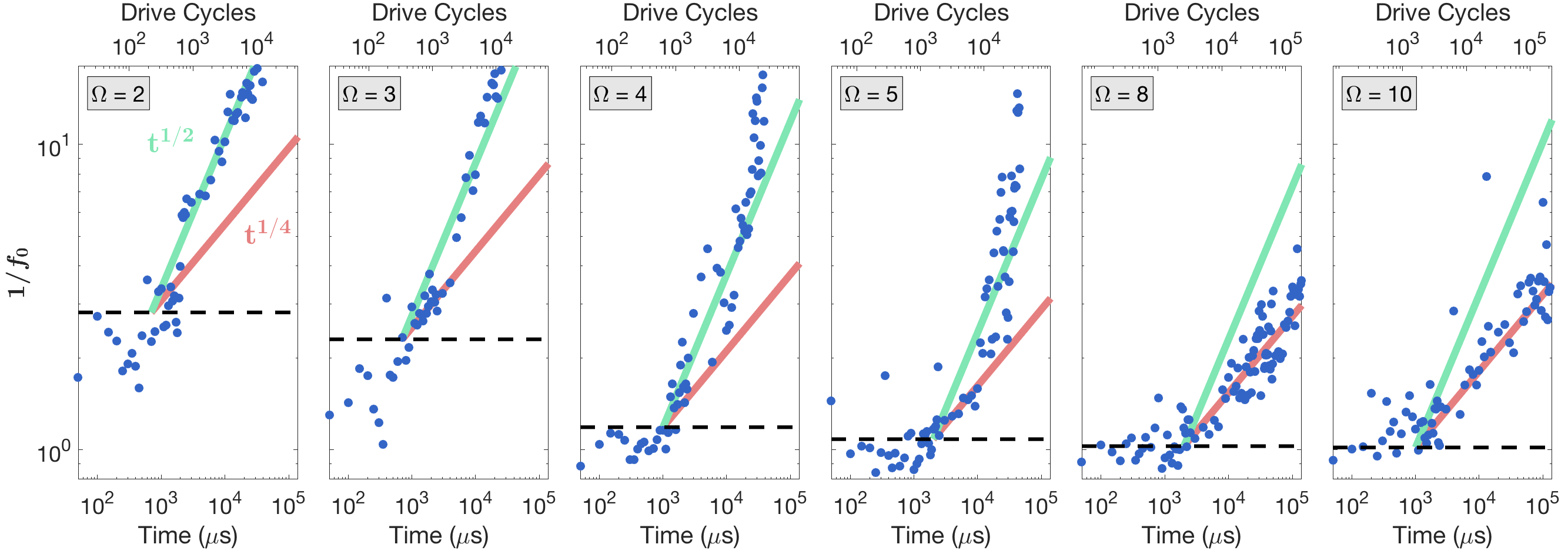}
\caption{Effect of drive frequency on the long-time evolution of Floquet matter. Panels show evolution of the participation ratio $1/f_0$ for $a=30$~nm, $\alpha=3$, and varying drive frequency $\Omega$ as indicated in the inset. Dashed line indicates the PGE prediction for the constant-$f_0$ plateau which extends as the frequency increases. The time-dependence of heating away from the prethermal plateau is observed to be strongly $\Omega$-dependent: at low frequency the participation ratio grows approximately as the square root of time (green solid line), while at the highest frequency the heating is consistent with a $t^{1/4}$ dependence (red solid line).}
\label{fig:timetrace}
\end{figure*}

To probe the effects of varying interparticle interactions, we measure the evolution of strongly-driven Floquet matter over much longer timescales than those shown in Figs.~2~and~3.  The long-time evolution of interacting driven systems is both especially relevant for the realization of useful many-body Floquet engineering, and especially challenging to address theoretically. Fig.~4A compares the initial evolution of samples with different Feshbach-tuned interaction strengths under the same drive parameters ($\Omega\!=\!2.6$, $\alpha\!=\!3$). For all three values of the interaction strength, the early-time dynamics are in agreement with the PGE description. As the system continues to evolve to large times over thousands of drive cycles, the PGE plateau decays. Strikingly, for a wide range of parameters we observe, as shown in Fig.~4B, that the system enters a second plateau in which $f_0$ is non-zero but smaller than in the first plateau.  Non-interacting and weakly-interacting samples remain in this second plateau for at least twenty thousand drive cycles at $\Omega=2.6$. We hypothesize that this second plateau can be understood as a consequence of a slow spreading in both position and quasimomentum: one relevant effect among others is that increasing quasimomentum extent will break the selection rules that prohibit odd-band occupation at $k=0$. To account in part for this spreading, all measurements of $f_0$ in Figs.~4~and~5 integrate over the entirety of the first Brillouin zone. The strongest-interacting samples, subjected to the same drive, behave in a fundamentally different way: they do not exhibit this second plateau but instead are observed to attain ergodicity, heating up indefinitely with no detectable atoms in the ground band after modulation. The second main result of this report, after the maps of the IPR and drive-dependent material properties, comprises these measurements of the detailed time evolution of Floquet matter: at short time scales, we observe a rapid emergence of a PGE prethermal state exhibiting large fluctuations, and at long time scales, we observe a second prethermal plateau and interaction-dependent ergodicity.
 
These results raise a crucial question: is the rapid onset of ergodicity for the strongest-interacting system inevitable? To quantitatively explore the possibility of delaying the onset of ergodicity in driven interacting systems, we measured the long-time evolution of the participation ratio $1/f_0$ for the strongest-interacting samples at \mbox{$\alpha\!=\!3$} and increasing values of $\Omega$. The results are shown in Fig.~5.   At \mbox{$\Omega\!=\!2$}, the observed long-time evolution of $1/f_0$ is consistent with the $\sqrt{t}$ dependence naively expected from Joule heating, and no significant plateau is observed. As the drive frequency is increased holding $\alpha$ constant, we observe the emergence of a quasi-static prethermal plateau lasting thousands of drive cycles, at a value of $f_0$ consistent with the PGE prediction.  This plateau too eventually decays. The lifetime of the high-frequency plateau is approximately ten times the timescale associated with the initial Thomas-Fermi interaction energies. We note that this high-frequency stabilization of the prethermal plateau cannot be simply explained by the exponential suppression of Floquet heating, predicted for single-band and spin models~\cite{Abanin:2015aa,KUWAHARA201696,huveneers,chetanNorm}, though it may be related; neither is the prethermal state characterized by simple high-frequency Floquet-Magnus-type expansions.  

Intriguingly, for the highest-frequency drives, the long-time departure from this plateau is significantly slower than for the lower drive frequencies, and clearly rules out the $\sqrt{t}$ dependence expected for Joule heating. In this regime $1/f_0$ obeys an approximate $t^{1/4}$ time dependence after departure from the prethermal plateau. This unusual behavior, while still poorly understood, is consistent with the sub-Joule heating predicted by recent related theoretical analyses~\cite{chandranfloquet,Weidinger:2017aa}.  Future work measuring the dynamics of heating away from the prethermal plateau in other regimes of drive strength, interaction strength, and modulation type should further elucidate the generality of these  predictions and experimental observations. The recovery of the prethermal plateau by increasing the drive frequency in the presence of interactions, and the observation of anomalously slow heating dynamics away from the recovered plateau, together constitute the third and final main result of this report.


In conclusion, we have used a flexible platform for studying strongly driven interacting quantum systems to acquire a complete prethermal map of the drive-dependent properties of tunable Floquet matter, revealing a Floquet delocalization crossover. The quantitative agreement between measured data and theoretical calculations provides an experimental confirmation that the prethermal state is describable by a periodic Gibbs ensemble. The measurement of an IPR via a double quench protocol introduced here represents a powerful quantitative tool for characterizing strongly-driven quantum systems in a variety of experimental contexts. Measuring the evolution of driven ensembles at both short and long times, we have observed two prethermal plateaux and a long-time transition to ergodicity at a rate and onset time which depend critically on drive frequency and interaction strength. 

\begin{acknowledgments}
The authors thank J\"org Schmiedmayer, Tarun Grover, Norman Yao, Chetan Nayak, Sid Parameswaran, Sarang Gopalakrishnan, Chaitanya Murthy, and Michael Knap for useful discussions, thank Sean Frazier for experimental assistance, and acknowledge support from the Army Research Office (PECASE W911NF1410154 and MURI W911NF1710323), the National Science Foundation (CAREER 1555313), the Office of Naval Research (N00014-14-1-0805 and N00014-16-1-2225), and the German Research Foundation (DFG) via the Research Unit FOR 2414 and the Gottfried Wilhelm Leibniz Prize program.  DW thanks Eva Lindroth and Stockholm University for hospitality during the preparation of the manuscript.
\end{acknowledgments}

K.S. and C.J.F. contributed equally to this work.

\appendix

\section{Sample preparation and loading}

The experiments begin with evaporation of bosonic ${}^7$Li atoms in a crossed optical dipole trap ($\lambda$=1064~nm, 7~W per beam, 100~$\mu$m beam waist, 1~kHz transverse trap frequencies) to generate a Bose-Einstein Condensate (BEC) with approximately 100,000 atoms at a temperature of 20~nK. During and after evaporation, the interatomic interaction strength is controlled by Feshbach tuning using a homogeneous magnetic field. We hold the BEC in the optical trap while ramping the field from the value used for evaporation (729~G, $a\!\approx$30~nm) 
to the desired final value in 100~ms. We then adiabatically load the atoms into the ground state of a combined 1D optical lattice with an initial static lattice depth of $V_{0} = 10 E_{R}$, where $E_{R}= \frac{\hbar^{2}k_{L}^{2}}{2m}$ is the lattice recoil energy. In the static lattice, the tunneling rate between lattice sites is 483 Hz and the lattice site trap frequency is $\omega_{0}$ = 159 kHz. The transverse confinement is provided by the Gaussian lattice beams, resulting in a transverse trapping frequency of 449~Hz$\times\sqrt{1+\alpha/2}$. We observe no significant excitation of transverse oscillator modes in the experiments reported here. Any additional forces along the lattice direction arising from magnetic field curvature or lattice beam intensity gradients are nulled out using magnetic shim coils to increase the period of Bloch oscillations to time scales significantly longer than our longest experiments~\cite{RSBO-PRL}. The interacting experiments are performed at Feshbach-induced {\it s}-wave scattering lengths of 2 nm and 30 nm, resulting in Thomas-Fermi interaction energies of 1.3 kHz and 3.8 kHz, respectively, for $\alpha=0$. Because the Thomas-Fermi interaction energies grow weakly with the modulation depth due to increased transverse confinement, we characterize interactions by the scattering length, which does not depend on the drive parameters.
 
 \section{Optical lattice for sign-changing modulation}
 
To enable realization of the $\alpha>1$ regime of ultrastrong lattice modulation, a combined lattice is formed by overlapping two 1D optical lattices with a relative spatial phase shift of half a period.  We use up to 7~W of 1064~nm light per beam and an 88~$\mu$m beam waist. The two lattices are separated in frequency by 160~MHz and have orthogonal linear polarizations. The beams are retroreflected by the same mirror to form two independent lattices. The relative phase is controlled by means of a waveplate stack, arranged so that the one of the lattices receives a $\lambda$/4 phase shift as it is retroreflected. This causes the two lattices to cancel each other when both beams have the same power, resulting in a featureless optical dipole trap; intentionally imbalancing the power results in a lattice of controllable sign. The depth of the combined lattice and the relative spatial phase between the two lattices are calibrated using matter-wave diffraction. After ramp-up of the combined lattice to an initial depth of 10~$E_{R}$, the system is quenched into the Floquet Hamiltonian by applying lattice amplitude modulation with some $\Omega$ and $\alpha$. The amplitude of the combined lattice is modulated at up to 2~MHz by simultaneously varying the power of RF signals sent to two acousto-optical modulators from an AD9854 DDS board. Crucially for the results we present, this double lattice modulation allows us to create a combined optical lattice that can change sign, where maxima (minima) become minima (maxima) during a drive cycle. 

\section{Measurement protocol}

After the system is allowed to evolve for some time in the modulated lattice, the modulation is quenched off and the combined lattice can either be snapped off or ramped off adiabatically with respect to the bandgaps to perform band-mapping ~\cite{bandmapping}. When measuring a prethermal map of Floquet material properties at a fixed total drive time like those shown in Figs.~1B~and~1C, the drive is allowed to fully complete the final modulation cycle before band-mapping, but for following the full time-dependent evolution as is done in Figs.~2C~and~3A the quench can be performed at any point in the drive cycle.  After band-mapping, resonant absorption imaging measures the atom number in the ground and excited bands. All band-mapping measurements are performed at a time of flight of 1.25 ms. At this time of flight, convolution of the initial Heisenberg-limited spatial distribution of the condensate with the momentum distribution limits our quasimomentum resolution to $\sim 0.2k_{L}$; this motivates our integration over the central 40\% of the Brillouin zone for the data shown in Figs. 1, 2, and 3. While population is initially concentrated in the even bands due to the even parity of the drive, odd-band population can result from experimental imperfections or from the weakening of parity-based selection rules away from zero quasimomentum.
 
\section{Identification of $f_0$ as an IPR for quantifying ergodicity}
\label{IPRappendix}

Here we show explicitly that for a non-interacting driven system the ground-band occupation $f_0$ is equal to the inverse participation ratio in the Floquet state basis. Furthermore we argue that for interacting systems in the prethermal regime, where the PGE can be applied, $f_0$ provides a straightforward and experimentally accessible metric for localization and ergodicity. 
 
The time-dependent Schr\"odinger equation of a driven system possesses quasi-stationary solutions called Floquet states, which are of the form $|n(t)\rangle e^{it\varepsilon_n/\hbar}$, with real quasienergy $\varepsilon_n$ and time-periodic Floquet mode $|n(t)\rangle=|n(t+T)\rangle$, where $T=2\pi/\omega$ denotes the driving period. For each time $t$ they form an orthogonal basis, so that for a given pure initial state $|\psi(0)\rangle$ the evolved state can be expressed as $|\psi(t)\rangle=\sum_n c_n |n(t)\rangle e^{-i\varepsilon_n t/\hbar}$, with time-independent coefficients $c_n=\langle n(0)|\psi(0)\rangle$. Accordingly, the expectation value of an observable $\hat{O}$ evolves as $\langle\hat{O}\rangle (t)=\sum_{nn'} c_n^* c_{n'}\langle n(t)|\hat{O}|n'(t)\rangle e^{it(\varepsilon_n-\varepsilon_n')/\hbar}$. The relaxation to a quasi-steady state (i.e.\ a time-periodic state) in the long-time limit can be associated with the dephasing and averaging out of the off-diagonal terms, so that asymptotically $\langle\hat{O}\rangle (t)\simeq \sum_{n} |c_n|^2 \langle n(t)|\hat{O}|n(t)\rangle$, corresponding to a Floquet diagonal ensemble described by a periodic density operator $\hat{\rho}_{\mathrm{dia}}(t)=\sum_n |c_n|^2|n(t)\rangle\langle n(t)|$~\cite{PRL257201}. For a non-interacting driven gas, there are an extensive number of integrals of motion given by the number operators $\hat{n}_j(t)$ of the single-particle Floquet modes $|j(t)\rangle=|j(t+T)\rangle$. The expectation values of these operators determine the PGE. To quantify the degree of localization (non-ergodicity), we initially prepare the system in the undriven ground-state $|\psi_0\rangle$, so that $c_n=\langle n(0)|\psi_0\rangle$, and consider the expectation value of the projector $\hat{O}=|\psi_0\rangle\langle\psi_0|$ at stroboscopic times $t_\nu=\nu T$ with integer $\nu$, which is equal to the squared overlap  $|\langle\psi_0|\psi(t_\nu)\rangle|^2$ with the evolved state $|\psi(t_\nu)\rangle$. According to the diagonal ensemble and employing $|n(t_\nu)\rangle = |n(0)\rangle$, the long-time average (indicated by an overbar) over the stroboscopic dynamics of this quantity gives 
\begin{equation}
\overline{{|\langle \psi_0 |\psi(t_\nu)\rangle|^2 }}
= \sum_n|\langle\psi_0|n(0)\rangle|^4 \equiv \mathrm{IPR}.
\end{equation}
This quantity is directly identifiable as the inverse participation 
ratio that quantifies the localization of the ground state
$|\psi_0\rangle$ in the basis of the Floquet states. Its inverse, $1/\mathrm{IPR}$, measures the number of Floquet states required to represent 
$|\psi_0\rangle$.
For the non-interacting gas, we find that the desired overlap is given by the fraction 
of atoms populating the single-particle ground state (i.e.\ the quasimomentum $k=0$ mode in the 
lowest Bloch band), $|\langle \psi_0 |\psi(t_\nu)\rangle|^2 =f_0(t_\nu)$, which we have measured for example in Fig.~1B. Thus, having relaxed to a quasi-steady state, we have
\begin{equation}
\mathrm{IPR} = \overline{f_0(t_\nu)},
\end{equation}
so that the measured observable directly quantifies non-ergodicity. 

In an interacting many-body system, the diagonal ensemble is still formally characterized by an exponentially large number of probabilities $|c_n|^2$. However, it is believed that the quasi-steady state is characterized by a periodic Gibbs ensemble (PGE), $\rho_{PGE}(t)=Z^{-1}\exp[-\sum_j\lambda_j\hat{I}_j(t)]$, with $\hat{I}_j(t)=\hat{I}_j(t+T)$ denoting the integrals of motion of the system~\cite{Lazarides:2014aa}. It is important to note that while a generic interacting Floquet system should at sufficiently long times approach a fully ergodic high-temperature state $\rho_{\mathrm{ergodic}}\propto 1 $~\cite{Lazarides:2014ab,PRX041048}, even the interacting system can approach a prethermal state that is accurately described by the PGE on intermediate, and potentially exponentially long, time scales.  This motivates the use of $f_0$ as a prethermal diagnostic of Floquet localization even in the interacting regime.

\section{Calculating expected number of occupied bands}

\label{finitenum}

Here we briefly discuss the reason that only a finite number of bands are expected to be significantly occupied even in the presence of a strong drive. Since quasimomentum is conserved in the absence of interactions, the single-particle Hamiltonian $H$ of the system can be expressed in terms of states $|m\rangle$ having momentum wave numbers $K_m = (4\pi/\lambda)m$ for integer $m$. In units of the recoil energy $E_R$, 
\begin{equation}
H = 4 \sum_m \bigg[ m^{2} |m\rangle \langle m| + \frac{V(t)}{16} \bigg(|m+1 \rangle \langle m|+|m-1\rangle \langle m| \bigg)  \bigg] \label{fewstates}
\end{equation}
with $V(t)=(V_0/E_{R}) [1+\alpha \cos(\omega t)]$.
Thus, already for $|m|\simeq 1$, the energy separation $(m+1)^{2}-m^{2} =2m+1$ to higher-lying states becomes larger than the time-dependent coupling matrix elements $V(t)/16$ and the driving frequency. This suggests that the drive will cause substantial redistribution among small $m$, while for large $m$ the Floquet states will not significantly differ from the undriven eigenstates of the system that correspond to the scattering continuum. In other words, except for very narrow resonances requiring both fine tuning and long times to make themselves felt, the ground state of the non-interacting system will mainly overlap with a few Floquet states.  See Ref.~\cite{HoneKohnFloquet} for a rigorous treatment of this problem.

The Hamiltonian of Eq.~(\ref{fewstates}) explains also the origin of the crossover to a highly localized regime with $f_0$ close to one for large frequencies, as we observe in Fig.~1. The fact that it couples only neighboring momenta $m$ and $m\pm1$ shows that (except for narrow resonances) the ground state is predominantly coupled to low energy states, transitions to which become off-resonant for large drive frequencies.  

%
%
%
%
\end{document}